# Visions of Revolutions:

# Microphysics and Cosmophysics in the 1930s[*]

Helge Kragh

Centre for Science Studies, Aarhus University, Denmark

**Abstract:** By 1930, at a time when the new physics based on relativity and quantum theory had reached a state of consolidation, problems of a foundational kind began to abound. Physicists began to speak of a new "crisis" and envisage a forthcoming "revolution" of a scale similar to the one in the mid-1920s. The perceived crisis was an issue not only in microphysics but also in cosmology, where it resulted in ambitious cosmophysical theories that transcended the ordinary methods of physics. The uncertain cognitive situation was, in some circles, associated to the uncertain political and moral situation. Did the problems of foundational physics demand a revolution in thinking that somehow paralleled the political revolutions of the time? I argue that although such ideas were indeed discussed in the 1930s, they were more rhetoric than reality. With the benefit of hindsight one can see that the perceived crisis was only temporary and not significantly related to social or ideological developments in the decade.

## 1. Introduction

In this paper I focus on aspects of the physical sciences in the 1930s, belonging to two different areas which were both of a foundational nature but the connections between which were contested. I argue that there was, at least in some quarters, a sense of crisis and a willingness to explore new approaches to

---

fundamental physics, to question old truths and to look for new ones in areas hitherto unexplored. A minority of physicists and astronomers suggested ideas and methods which by the standards of the time (and indeed by most later standards) appeared to be rather extreme. Some of these ideas belonged to microphysics, others to the physics of the cosmos which for the first time was seen as deeply connected with, and perhaps even determining, the ordinary laws of laboratory physics. Cosmology was a new science, and its meaning and relation to other sciences was a disputed issue. Most of the paper is a non-technical survey of themes in fundamental physics in the period from about 1928, when quantum mechanics was largely consolidated, until the end of the following decade.

In the last two sections I also look briefly at possible connections between the political and ideological climate of the decade, characterized as it was by an increasing polarization between authoritarian ideologies, and the developments in physics. "Crisis" was a common theme in both areas, which may suggest a causal connection between the trends in politics and ideology and what happened in the physical sciences. However, tempting as such a suggestion may be, I conclude that its lacks documentary evidence.

In an acute analysis of the early development of modern theoretical physics, Suman Seth has argued that as a historical and descriptive category, the term "crisis" should be used as an actors' category (Seth 2007). It follows that the actors – in this case the physicists – need to express an awareness of crisis and refer to the situation with this or corresponding terms. This is indeed what I have in mind when I speak about crisis in the 1930s: many physicists, if far from all, perceived their science to be in a state of crisis and consequently looked for new approaches to the problems of fundamental physics. That this perceived crisis was, in a sense, an illusion, was only realized after World War II. As seen from the perspective of the 1930s, it was real enough.

## 2. The perceived crisis in microphysics

By the late 1920s the revolution in physics was largely over – or so it may seem in retrospect. Einstein's special theory of relativity was accepted as an unproblematic and necessary tool, and although few physicists cared about the general theory of relativity, it, too, enjoyed general acceptance. The problem with general relativity was not that it might be wrong, but rather that it was seen as irrelevant to mainstream physics, whether experimental or theoretical. The theory was cultivated by a small group of mathematicians and mathematically minded physicists and astronomers, without the majority of physicists paying much attention. As far as applications were concerned, these were restricted to areas of astronomy and cosmology. This period of low-water mark of general relativity lasted throughout the 1930s and 1940s (Eisenstaedt 1989).

Quantum physics proceeded in an entirely different manner. The new theory of quantum mechanics had developed with unprecedented speed into a theory of matter and radiation which celebrated one triumph after the other. The year of 1928 witnessed not only Paul Dirac's fundamental wave equation of the electron, but also George Gamow's wave-mechanical explanation of alpha radioactivity (suggested independently by Ronald Gurney and Edward Condon) which demonstrated that even the mysterious atomic nucleus was unable to resist quantum-mechanical analysis.

In the introduction to a paper of 1929 on many-electron atoms, Dirac looked back on the stormy development of quantum physics. He expressed the view, at the same time optimistic and reductionistic, that apart from relativistic effects quantum theory was essentially complete. In an oft-quoted passage he wrote: "The general theory of quantum mechanics is now almost complete. … The underlying physical laws necessary for the mathematical theory of a large part of physics and the whole of chemistry are thus completely known, and the difficulty is only that the exact application of these laws leads to equations much too complicated to be soluble" (Dirac 1929: 714; Simões 2002). Dirac was not the only one to feel this way. Léon Rosenfeld recalled conversations from about the same time when he was told that within a few years, "physics will be finished." After that, "We shall then turn to biology" (Rosenfeld 1967: 114).

Yet, at about the same time clouds began to gather. Investigations of the spectrum of beta radioactivity seemed to indicate a violation of energy conservation; understanding the cosmic rays in terms of quantum mechanics proved to be frustratingly difficult; and attempts to formulate a unified theory of electromagnetism and quantum mechanics – known as quantum electrodynamics or QED – led to so-called divergencies, that is, infinite (and hence unphysical) quantities. In addition, how was one to explain the energy production of the stars? And what about the mysterious negative-energy solutions turning up in the relativistic Dirac equation? Particles cannot have negative energy, and yet the equation apparently predicted such monstrous entities. The point is that at a time when quantum physics was assumed to have reached completion, problems suddenly began to abound (Kragh 1999: 196-201).

Some of the key players in theoretical physics were deeply worried and believed for a while that the paradoxes of nuclear physics required drastic changes in theory, perhaps some "new revolution" comparable to the one that recently had given rise to quantum mechanics. The Russian theorist Lev Landau suggested in 1932 that stars might possess high-density, so-called "pathological regions" in which the known laws of quantum physics failed and new were required (Landau 1932).

The search for new laws of physics might direct physicists towards the stars, but it might also direct them towards the equally mysterious phenomenon of life. The promising young physicist Max Delbrück was greatly inspired by Niels Bohr's 1932 address on "Light and Life," and his turn from physics to biology has generally been seen as motivated by the hope that the study of biological phenomena might lead to new laws of physics (Stent 1998).[1] Pascual Jordan, one of the founders of quantum mechanics, similarly thought that biology might be a source of new fundamental advances in physics. During the 1930s he not only engaged in a holistically oriented research programme concerning biology and quantum physics (Beyler 1996), he was also led to examine the relationship between quantum theory and cosmology within a similar perspective. A declared positivist, Jordan found in quantum mechanics a philosophical perspective that led to the overthrow of materialism and an opening towards religion and other forms of spiritual quest (Wise 1994; Kragh 2004: 175-185; Beyler 2007). As far as cosmology was concerned, he engaged in numerological reasoning in the style of Eddington, except that he maintained

---

[1] However, Daniel McKaughan has argued that this interpretation is unfounded (McKaughan 2005).

that his approach was in full agreement with the norms of positivist empiricism.[2]

Another case in point is energy conservation – in the generalized, relativistic sense of mass-energy conservation – which is almost a dogma in physics and about the last thing physicists will abandon. Yet for a while Bohr and a few other physicists seriously believed that energy conservation might not be strictly valid in all nuclear processes. The heretical idea was pursued for a couple of years and received with enthusiasm by some of his younger colleagues, including the Leningrad physicists Gamow, Landau, and Matvei Bronstein. Only after the discovery of the neutron and Wolfgang Pauli's introduction of the neutrino as a saving device in beta decay did Bohr abandon his suggestion of violation of energy conservation. The idea was however revived by Dirac, who in 1936 suggested that relativistic field theory, including conservation of momentum and energy, should be replaced by a new kind of quantum theory in which energy was only statistically conserved (Dirac 1936).

As some physicists were willing to question energy conservation, so other physicists were willing to question another of the sacred principles of science, the immutability of natural laws (see below). And still others, including Werner Heisenberg, suggested to abandon the homogeneity of space-time and replace this standard assumption with a theory where space and time were composed of discrete units given by a smallest length and a smallest time interval (Kragh and Carazza 1994; Kragh 1996a). Proposals of this type, sometimes in rather bizarre versions, were entertained by several physicists in the period ca. 1928-1934 when they were discussed by Henry Flint, Arthur Ruark, Louis de Broglie, Gleb Wataghin, Dmitri Iwanenko, Pauli, and others. Although many of the suggestions of abandoning continuous space-time were speculative, they appealed to some of the leading quantum theorists, including Heisenberg, Pauli, and Erwin Schrödinger. When confronted with the grave problems facing quantum mechanics, they were quite willing to examine wild ideas. Einstein, too, speculated in the 1930s about a possible new physics based on discontinuous space-time. However, unable to see how such a view of physics could be formulated without ruining the entire edifice of existing theory, he shelved the project (Stachel 1993).

---

[2]  As pointed out in Wise (1994), Jordan's views on the consequences of the new physical world picture agree to some extent with Paul Forman's analysis of physics in the Weimar period (Forman 1971, reprinted pp. 85-202 in Carson et al. 2011). However, Wise also notes that there are important differences.

The ambitious quantum electrodynamics developed by Pauli and Heisenberg in 1929-30 had its own problems, in particular that a certain quantity known as the self-energy of the electron turned out to be infinite. And this was only one of the infinities that threatened the mathematical and conceptual structure of the new relativistic quantum theory and which was a source of much worry (Rueger 1992). Related problems arose in the attempt to understand the high-energy component of the cosmic radiation, which caused Hans Bethe, Walter Heitler and others to conclude that quantum mechanics broke down at high energies. As J. Robert Oppenheimer concisely summarized the situation in a letter of 1934, "theoretical physics – what with the haunting ghosts of neutrinos, the Copenhagen conviction, against all evidence, that cosmic rays are protons, Born's absolutely unquantizable field theory, the divergence difficulties with the positron, and the utter impossibility of making a rigorous calculation at all – is in a hell of a way" (Kragh 1999: 199). Heisenberg similarly wrote to Pauli in 1935: "With regard to quantum electrodynamics we are still in the state we were in 1922 concerning quantum mechanics. We know that everything is wrong" (Pauli 1985: 386).

The responses to what was perceived as a continued crisis varied, with a considerable part of the community of theoretical physicists, including Bohr, Dirac, Heisenberg, Pauli, and Landau, favouring a revolutionary approach. That is, they believed that the problems could not be solved within existing theory, or some modification of it, but should be exploited in constructing a theory of the future that might differ as much from existing quantum theory as this theory differed from classical theory.

The crisis in quantum theory was connected with and reinforced by the confusing situation that followed the breakdown of the so-called two-particle paradigm. Until 1930 it was commonly assumed that all matter consists of only two elementary particles, the electron and the proton, but with the introduction of new hypothetical particles (the neutrino, the positron and the magnetic monopole) the paradigm began to dissolve. Within a few years the neutron and the positron had become real particles, and still new particles with mass intermediate between the electron and the proton appeared to be hidden in the cosmic rays. What were these particles and how did they relate to theory? The "mesotron" detected in 1937 proved particularly troublesome and difficult to understand theoretically. The situation in the new science of elementary particle physics remained messy and complicated throughout the 1930s and the following decade.

It turned out that the crisis in microphysics was only temporary, as most of the problems could be solved within the established framework of physics.

What matters is that many physicists in the period expressed a revolutionary attitude in the sense that they more or less expected that some kind of new physics would follow in the near future. The unknown nature of the new physics only stimulated their willingness to consider very radical, sometimes extreme ideas. On the other hand, none of the physicists envisaged a revolution in the strong sense that Thomas Kuhn suggested in 1962, that is, a new physics incommensurable with the established one. It was generally agreed that the recently completed quantum mechanics had come to stay, if perhaps only as an approximation to the hypothetical quantum physics of the future. The advocates of crisis and revolutionary changes in physics did not form a homogeneous group, nor did they represent the majority of physicists. On the other hand, they included some of the leaders of the field, and to write them off as just a few isolated figures would be to underrate the persistence and significance of the feelings of crisis in the period.

### 3. The new cosmology.

Relativistic cosmology, based on Einstein's field equations of 1917, revolutionized the science of the universe. These equations included a new constant of nature, soon known as the "cosmological constant," that corresponded to a repulsive cosmic force but whose nature was uncertain and contested. Even more than quantum mechanics in the realm of microphysics, Einstein's theory established a new conceptual framework for the study of the universe that had no counterpart in earlier cosmological theories based on Newtonian theory (despite the development in the 1930s of Newtonian cosmologies analogous to the Einsteinian models). The revolutionary nature of the new cosmology was clear to its few practitioners at an early date, but without arousing much interest in the broader scientific community. For a long time cosmology was a somewhat esoteric field of research, cultivated only by a small minority of astronomers and physicists. The new microphysics based on quantum theory seemed to have nothing in common with the relativistic models of the universe, which was one reason for the lack of popularity of cosmology among physicists in the 1920s. However, the separation was not complete, and a few physicists suggested links between the two fields even at this early date.

An interesting attempt, which appears to have been overlooked by historians of science, was made by the Hungarian-German physicist Cornelius Lanczos in 1925, shortly before the advent of Heisenberg's quantum mechanics. Lanczos investigated an Einstein-like, matter-filled world model in which space varied periodically in time. In connection with the propagation of waves in his "spherical ring model," he was led to introduce quantum theory. For the world period he arrived at the expression $T = 4\pi^2 mR^2/h^2$, where $m$ denotes the mass of the electron and $h$ is Planck's constant. With a world radius $R$ of one million light years, this gives the enormous number $T \cong 10^{41}$ years. According to Lanczos, the quantum nature of microphysical phenomena reflected the state of the cosmos, rather than being a feature specific to the atomic world. "The solution of the quantum secrets are hidden in the spatial and temporal closedness of the world," he wrote (Lanczos 1925: 80).

The speculations of Lanczos and and a few others can be seen as precursors of the research programme that Arthur Eddington started developing in 1929 (and of which more below). It is also worth pointing out that the first version of the Big-Bang universe, proposed by the Belgian astrophysicist Georges Lemaître in 1931, was partly inspired by recent developments in quantum theory (Kragh and Lambert 2007). Lemaître's formulation reveals a close similarity to contemporary discussions concerning the validity of the space-time continuum in the quantum domain. His note of 1931 was probably indebted to the views of Bohr, who had recently argued that the usual concepts of space and time have only statistical validity and must be modified in the atomic domain. In general, Lemaître's hypothesis of a "primeval atom" as the beginning of the universe reflected the state of uncertainty and crisis in contemporary quantum physics.

The introduction of the expanding universe opened up for new possibilities and speculations, and in general helped to draw cosmology and microphysics more closely together. If the universe is not static, but in a state of continual expansion, it might suggest that perhaps the laws of nature, the very foundation of science, are not static either. It is in this period, the early 1930s, that we find the first ideas of laws and constants of nature varying in cosmic time. According to the process philosophy of Alfred North Whitehead, the British mathematician and philosopher, everything was in a state of flux, including the laws of nature. In 1933 he concluded that "The modern evolutionary view of the physical Universe should conceive of the laws of nature as evolving concurrently with the things constituting the environment" (Whitehead 1933: 143). A somewhat similar idea, but more concrete and less philosophical, was entertained by Bronstein, who the same year applied Bohr's vague idea of energy nonconservation to the entire universe and, moreover, suggested that Einstein's cosmological constant was not a true constant but that it decreased with time.

The radical idea of varying constants of nature, as it turned up in the 1930s, could be associated with the expanding universe (as Bronstein and Dirac did), but it could also be used to avoid this conclusion and save the static universe which many scientists continued to favour over the new cosmology. For example, by assuming the velocity of light $c$ or Planck's constant $h$ to vary in time, it was possible to account for the galactic redshifts and Hubble's empirical law without being forced to accept the expanding universe. Thus, P. Wold, a physicist at Union College, New York, suggested in 1935 that the speed of light might decrease slowly in time – he considered the linear variation $c = c_0(1 - kt)$ – and that this would lead to a redshift in agreement with Hubble's redshift-distance relation. There were several such proposals in the 1930s, but they were decidedly heterodox and failed to attract much attention from leading physicists and astronomers. Models based on the controversial hypothesis of a varying speed of light were eventually developed into quantitative theories of the early universe, but that only occurred much later (Kragh 2006).

The decade also witnessed a revival of opposition to the "heat death" scenario, the gloomy prediction from the second law of thermodynamics that in the far future the universe will end irreversibly in a lifeless soup of maximum-entropy matter and radiation. The idea of a unidirectional "degradation" of heat into useless heat went back to the 1860s, and with the discovery of another unidirectional process, the expansion of the universe, it seemed to be reinforced. Some scientists found the scenario to be intolerable from an emotional and intellectual point of view, and they consequently came up with alternatives to it. Reputed physical scientists, such as Nobel laureates Robert Millikan from the United States and Walther Nernst from Germany, proposed stationary universe models which had in common that they contradicted the second law and introduced hypotheses at variance with established physics, among which were the formation of matter out of starlight (Kragh 1996b: 143-161).

## 4. Cosmophysics

The willingness to explore new approaches to fundamental physics, and to establish links between microphysics and cosmology, was mostly a phenomenon that flourished in Great Britain, where a tradition of "cosmophysics" emerged in the 1930s, only to disappear rather abruptly after World War II (Kragh 1982). The most prominent of the new-style cosmophysicists were Eddington and E. Arthur Milne, but also Dirac, James Jeans, Edmund Whittaker and several others can reasonably be counted to the loose group of British cosmophysicists. In spite of adopting very different world models, Eddington and Milne had much in common when it came to methodology and the general aim of physics. Both argued for *a priori* principles from which the laws of nature were deducible by rational reasoning. Although they claimed to base their world systems on common sense experiences, in reality cosmophysics was characterized by a full-blown rationalism where experimental tests were of no more significance than in mathematical proofs (Kragh 2011: 87-116).

Eddington's overall research programme from about 1930 to his death in 1944 was to bridge relativistic cosmology with atomic and quantum theory, and to do it in such a way that *a priori* knowledge of nature could be obtained epistemologically, that is, by pure deductions from the peculiarities of the human mind. As he wrote, "it should be possible to judge whether the mathematical treatment and solutions are correct, without turning up the answer in the book of nature" (Eddington 1936: 3). Milne was no less convinced of the power of pure reasoning. He proudly declared that world physics was as much a matter of logic and reason as of observation and experiment (Milne 1935: 266):

> The philosopher may take comfort from the fact that, in spite of the much vaunted sway and dominance of pure observation and experiment in ordinary physics, world-physics propounds questions of an objective, non-metaphysical character which cannot be answered by observation but must be answered, if at all, by pure reason; natural philosophy is something bigger than the totality of conceivable observations.

Eddington claimed that he was able to deduce the numerical values of the fundamental, dimensionless constants of nature purely theoretically, and that they expressed some deep-lying connection between the atom and the cosmos. For example, he famously deduced that the inverse fine-structure constant, a measure of the strength of electromagnetic interactions, had the accurate value of $\alpha = hc/2\pi e^2 = 137$. He similarly used his theory to calculate Hubble's recession constant from the values of other constants, and found in this way a value in reasonable agreement with the one obtained from observations. Eddington's view of quantum mechanics was highly unorthodox, as he claimed that the theory necessarily implied considerations of the universe as a whole. The mass of an electron was conceived as an interchange energy with all the other charges in the universe, and hence of cosmological nature. Such ideas were in provoking contrast to the standard view concerning physical science, and consequently very few scientists wanted to associate themselves completely with his theory. Nonetheless, it was quite influential, both in science, philosophy, and in cultural circles.

At least a dozen physicists, both in Britain and on the Continent, were inspired by Eddington's research programme which they applied in various ways, more often than not in speculative attempts to find significant relationships between the constants of nature. Arthur Haas in Austria and Reinhold Fürth and Hans Ertel in Germany were among the most industrious of the Eddington epigones. Schrödinger was the only leading physicist of the 1930s who wholeheartedly supported Eddington's approach, if only for a while. He accepted Eddington's general position that experimental results are determined by the state of the universe as a whole, and in a couple of works from the late 1930s he adopted with enthusiasm this line of thinking. Schrödinger had high hopes for an Eddingtonian revolution in physics, but after a couple of years he came to the conclusion that Eddington's theory was basically beyond comprehension, an individual enterprise not communicable in ordinary scientific language (Kragh 1982; Rueger 1988).

By adopting elements from both Eddington and Milne, Dirac suggested in 1937 a reformation of cosmology based on the large dimensionless numbers that can be constructed from the fundamental constants of nature. Dirac took it as a general principle that all such numbers are interconnected, from which he deduced that the gravitational constant must vary inversely with the cosmic era. Even more radically, he also argued from the "large number hypothesis" that the number of particles in the observable universe will increase with time, namely proportionally with the square of the epoch since the Big Bang, meaning that his theory violated conservation of matter-energy. Dirac's cosmological theory was problematic from both an empirical and methodological point of view, yet he remained convinced that the large number hypothesis was a fundamental principle of such power and beauty that it had to be right. As he explained in a lecture of 1939, it followed from the principle that the laws of nature were evolutionary, not fixed once and for all (Dirac 1939: 128). Dirac even went as far as to suggest that the laws of nature at a given cosmic time were not the same all places in the universe:

> At the beginning of time the laws of Nature were probably very different from what they are now. Thus we should consider the laws of Nature as continually changing with the epoch, instead of as holding uniformly throughout space-time. … As we already have the laws of nature depending on the epoch, we should expect them also to depend on position in space, in order to preserve the beautiful idea of the theory of relativity that there is a fundamental similarity between space and time.

Although Dirac's cosmological theory was not widely discussed, it was taken up by a few other physicists and astronomers who felt fascinated by the large number hypothesis and the possibility of a varying gravitational constant. Jordan was particularly enthusiastic. In the late 1930s he combined in his own way Dirac's theory with elements of Eddington's numerological reasoning, and for a period of nearly twenty years he advocated various versions of cosmology with a gravitational constant decreasing in time. Strangely, in regard of the undeniable hypothetical and speculative elements in Dirac's theory, the positivist Jordan was convinced that the theory was in agreement with the epistemological requirements of empiricism. He considered it to be an "empirical cosmology" based on numerical relationships that were "mere reformulations of experience, freed from hypotheses" (Jordan 1937: 515).

Philip Morowski, a philosopher of science and economic thought, has argued that the more recent interest in varying constants of nature reflects cultural trends. It is, he says, an "indication of the postmodernist mind-set" (Mirowski 1992). However, there seems to be nothing particularly postmodernist about it. The interest certainly existed in the 1930s, a decade with a cultural climate very different from the later one characterized as postmodernism.

## 5. The debate on the "modern Aristotelians"

The kind of rationalistic cosmophysics expounded in different versions by Eddington, Milne, Dirac and Jordan enjoyed considerable popularity in Britain, but of course it was strongly opposed by more empirically minded scientists and philosophers. The astrophysicist and philosopher of science Herbert Dingle was among the first to launch a counter-attack against the unbalanced methods of what he called the "modern Aristotelians" (Kragh 1982; Urani and Gale 1994). Dingle defended the traditional, empirical-inductivist view of science which he saw threatened by the over-ambitious and arrogant "pseudo-science of invertebrate cosmythology" that ignored sense observation and built on unverifiable principles such as Milne's cosmological principle and Dirac's large number principle. He objected to such general principles which he took to be *a* *priori*, nothing but "chimeras" that seduced the imagination and led scientists away from the only genuine path of science, which he identified with the methods of Galileo and Newton.

Dingle's vituperative and strongly worded attack of 1937 caused a heated and most interesting debate in *Nature* that engaged many of Britain's most prominent scientists, including Harold Jeffreys, John B. S. Haldane, Charles Darwin, and Gerald Whitrow, and of course the main contestants Milne, Eddington, Dirac, and Dingle. The controversy spilled over into *The Observatory*, the monthly review published by the Royal Astronomical Society, where the astronomer George McVittie expressed his and others' objections in emotional terms: "It is eventually borne in on the puzzled reader that Milne and [Arthur G.] Walker are not trying to understand Nature but rather are telling Nature what she ought to be. If Nature is recalcitrant and refuses to fall in with their pattern so much the worse for her" (McVittie 1940: 280).

What is so instructive by this debate is that we have here an open discussion among top scientists about the very norms and standards of science, about which methods are legitimate and which are not. Whereas Dingle and Jeffreys accused their opponents of sinning against positivistic virtues, and thereby betraying the very foundation of science, Milne and Eddington maintained that knowledge about nature could and should be derived from pure rational thinking. As Milne made clear, he aimed at a completely rational explanation that allowed no irreducible laws or brute facts. His ideal of science had much in common with the view Descartes had advocated 300 years earlier. He often likened the true laws of nature with geometrical theorems. After God had created the world and the laws of nature – and Milne was convinced of divine creation – the world developed strictly according to laws that were not at God's further disposal.[3]

What really alarmed Dingle and his allies were not only the extravagant theories of a few physicists but even more so the cultural and political standards they reflected. "My concern," he wrote, "is with the general intellectual miasma that threatens to envelop the world of science." In his view, cosmophysics was associated with authoritarianism, and in 1937 it was all too clear that such thinking was potentially dangerous (Dingle 1937):

> The times are not so auspicious that we can rest comfortably in a mental atmosphere in which the ideas fittest to survive are not those which stand in the most rational relation to experience, but those which can don the most impressive garb of pseudo-profundity. There is evidence enough on the Continent of the effects of doctrines derived "rationally without recourse to experience." To purify the air seems to me an urgent necessity: I wish it were in other and better hands.

In the case of Dingle and a few other physicists one can see in the 1930s connections of the type discussed by Paul Forman in his celebrated essay of 1971, namely that extrinsic influences led physicists to hope and search for a new understanding of nature. Dingle was certainly not among them, but he believed there were such worrying influences in the case of cosmophysics. However, his crusade was essentially his own and he merely issued his warning against excessive rationalism in general terms. In fact, it is difficult to find examples in either cosmophysics or quantum theory of physicists who responded to extrinsic, political-ideological influences in the manner suggested by Forman. Jordan may seem to be an example, but neither he nor other major figures in physics capitulated to a hostile cultural environment such as can be argued in the case of the Weimar Republic. As to Jordan, he rather led the way (Wise 1994: 227).

---

[3] Religion and spiritual attitudes played a considerable role in cosmology and fundamental physics in the 1930s, if more in the reception of the new physics than among the physicists themselves. For discussion and references, see Kragh (2004).

Dingle's reference to continental doctrines might have alluded to the situation in Nazi Germany, which at the time caused growing alarm among British intellectuals.[4] At any rate, Dingle's statement should be read in relation to the prevailing depreciation of science in the period. In Britain there was a concern about anti-science feelings, mysticism, and anti-intellectualism, attitudes which gained terrain and (so Dingle may have feared) might be stimulated by the kind of cosmophysics expounded by Milne, Dirac, and Eddington. In his pioneering work *The Social Function of Science*, published 1939, the Marxist crystallographer John D. Bernal may have alluded to cosmophysics when he wrote:

> This mysticism and abandonment of rational thought is not only a sign of popular and political disquiet; it penetrates far into the structure of science itself. The working scientist may repute it as firmly as ever, but scientific theories, particularly those metaphysical and mystical theories which touch on the universe at large or the nature of life, which had been laughed out of court in the eighteenth and nineteenth centuries, are attempting to win their way back into scientific acceptance. … In this way, modern science is being made an ally of ancient religion, and even to a large extent a substitute for it.

---

[4] If so, it is misleading. The fascist *Deutsche Physik* movement, as propagated by Phillipp Lenard, Johannes Stark and others, in fact boasted of being pragmatic, empiricist and anti-doctrinaire.

Bernal warned against these tendencies and referred explicitly to Jeans, Eddington, and Whitehead as exponents of "a new scientific mythical religion" (Bernal 1939: 3-5).

But not all red scientists felt deterred by the tendencies represented in cosmophysics. Like Bernal, the evolutionary biologist John B. S. Haldane was a member of the British Communist Party (he only became a full member in 1942). He had for some time followed the development in physics and cosmology, and came to sympathize with Milne's cosmological theory because of its evolutionary nature (Kragh 2004: 221-224). Haldane took the side of Milne and Dirac in the 1937 debate in *Nature*. What appealed to him was the idea that the laws of nature are historical phenomena, not eternal absolutes provided by God. Somehow he convinced himself that Milne's world physics was "beautifully dialectical" and in close harmony with Marxist thought. In his notes for the English translation of Friedrich Engels' *Dialectics of Nature* he even suggested that Engels would have welcomed Milne's theory. However, given that Engels, a radical atheist, was strongly opposed to all forms of non-eternal cosmologies, this is most unlikely.[5] At any rate, it was a view not shared by scientists and party philosophers in the Soviet Union, where Milne and Eddington were dismissed as idealists and reactionary apologetes.

---

[5] It was a central doctrine in Engels' philosophy of nature that matter and motion could neither be created nor destroyed. This he spelled out in his *Anti-Dühring* of 1878, and in his *Dialectics of Nature* he made it clear that he considered the universe as a whole to be eternally recurring (Engels 1940: 25).

Another left-wing response to the situation in physics came from the young Marxist writer Christopher Caudwell whose book *The Crisis in Physics* appeared posthumously in 1939, two years after he had been killed in Spain (Caudwell 1939; Thompson 1995). Caudwell was, like Bernal, concerned with the anti-materialistic tendencies which had led to "the crisis in bourgeois physics," and particularly with the popular expositions of people like Eddington and Jeans whom he criticized as symbols of "the mentalism and tendency towards anti-scientific scholasticism." He furthermore suggested that there was a causal connection between the crisis in physics and the moral and political crises in society. Realizing that the majority of physicists were neither revolutionaries nor socialists, he nonetheless suggested that "to adopt a genuine revolutionary standpoint in physical theory involves the adoption of a genuinely revolutionary attitude in real life." Contrary to some other Marxists, Caudwell did not offer a materialist alternative and seems to have thought that the illness was not so much in the physical theories themselves as in their idealistic interpretations. The crisis in physics reflected the existence of two conflicting ideologies in the late-capitalist era – scientific materialism and philosophical idealism (see also Cross 1991).

Tunes, somewhat similar to those of Bernal and Caudwell, could be heard from very different political positions. Thus Max Planck, a conservative in both politics and science, spoke in several addresses from the 1930s of the contemporary "moment of crisis" which had invaded physics. Planck emphatically rejected facile inferences from the formalism of quantum physics to problems in biology, psychology and philosophy, and in general he was worried over what he saw as a tendency towards extremism combined with declining standards of critical thinking in physics. "There is scarcely a scientific axiom that is not nowadays denied by somebody," he lamented in an address of 1933. "And at the same time almost any nonsensical theory that may be put forward in the name of science would be almost sure to find believers and disciples somewhere or other" (Planck 1958: 250-251; Heilbron 1968: 142-143).

For another voice of the period, listen to Viscount Samuel, who in a 1938 Royal Institution lecture discussed the relationship between science and philosophy (excerpted in *Nature* 143: 210). Europe was prepared for war, he observed, and behind the armies were the political creeds, such as Communism, National Socialism, fascism and democracy, some of which creeds were decidedly anti-intellectual. He suggested that there existed an unholy alliance between the forces of darkness and idealist philosophies of science, and urged a return to an empiricist and realist view of nature.

Max Born, who in 1934 had emigrated from Hitler's Germany and at the time was professor in Edinburgh, broadly agreed. He attacked Eddington's theory as an example of excessive rationalism and warned against physical theories based on abstract reason and principles of a cosmological nature. In letters to Einstein of 1944, he characterized the Milne-Eddington approach as "rubbish" and an example of Hegelianism in physics (Kragh 1982: 104). And yet, in spite of his dismissal of cosmophysics, Born was not unreceptive to the aspirations of Eddington and Milne, which had more than a little affinity with his own inclination toward ambitious unitary theories. During the late 1930s he was busy with developing his own unitary theory, which included calculation of the mysterious fine-structure constant $\alpha$. As he wrote in a paper of 1935, "A perfect theory should be able to derive the number $\alpha$ by purely mathematical reasoning without recourse to experiment" (Born 1935: 560). The statement could as well have been Eddington's, but Born did not notice the similarity.

## 6. Discussion

The well-known British historian Eric Hobsbawm published in 1994 a book entitled *The Age of Extremes*, a thematic world history from the start of World War I to the fall of the Soviet empire and the Gulf War (Hobsbawm 1994). Rather unusual for a book of its kind, it includes a substantial chapter on the natural sciences and refers explicitly to the "crisis in physics" in the 1930s. If the twentieth century was an extreme age from the perspective of politics and ideas, so it was from the perspective of science, although there have been periods more extreme than others. As far as the physical sciences are concerned, the interwar period may well qualify as an extreme era.

The 1930s was basically a decade of consolidation and progress in physics, a period when the new quantum theory was successfully applied to the stars, the solid state of matter, the aurora borealis, the structure of chemical compounds, and the atomic nucleus. But it is not this steady flow of successes I have dealt with in my survey. On the contrary, I have focused on the problems and frustrations that were *also* parts of the physics of the decade. Running parallel with the progress and the new discoveries there was in some quarters an undercurrent of dissatisfaction with the foundations of physics, a revolutionary spirit seeking for new laws and new ways of thinking. One of these ways was to integrate microphysics with cosmology, a trend which somehow resonated with the period's zeitgeist and was particularly strong among British scientists.

Although the attempts of creating new physics were diverse, and in general had little in common, they may be said to represent a revolt against narrow, positivistic science and an openness towards heterodox and extraordinary forms of physics, even towards metaphysics (and, in the case of Eddington, mysticism, such as analyzed in Stanley 2007). In this respect the intellectual climate of physics in the 1930s differed drastically from the climate in the first two or three decades after World War II.

The 1930s started with the great economic depression and ended with the outbreak of World War II. In between, the decade was characterized by the rise of despotic dictatorships and several "minor" wars such as Italy's war against Ethiopia, Japan's against China, and the Spanish Civil War. Physics and the other sciences were not unaffected by the fluctuations in political and ideological temperature, but by and large the influence was limited to the organizational, political and economic aspects of science. The political development did not affect the cognitive level of physics directly or to any great extent. There were indeed various claims in the period of such an influence, but little evidence that it actually occurred.

Whereas the ideological climate was undoubtedly a factor in the popular reception of the new physics and how it was interpreted by religious and other cultural groups, its impact on the more established theories of physics seems to have been very limited.[6] At any rate, I can see no common pattern between the fluctuations in politics and the physical sciences. Sure, they may both have been manifestations of a *Zeitgeist* characteristic of the period, but this is a concept which is too nebulous to serve as a useful explanatory link. Hobsbawm apparently was of the same opinion. "Can it be said that these fluctuations in political and ideological temperature affected the progress of the natural sciences?" he asks. His answer was essentially no. "While it is easy to show that the conflicting schools and changing fashions in economic thought directly reflect contemporary experience and ideological debate, this is not so in cosmology" (Hobsbawm 1994: 547-548).

Things may have looked different in the 1930s, at least if seen from a socialist perspective. Hyman Levy, an applied mathematician and spokesman of the red scientists' movement in Great Britain, wrote an introduction to *The Crisis in Physics* in which he essentially repeated Caudwell's thesis of an unconscious connection. Whether scientists were aware of it or not, "any scientific theory is necessarily the specialized development of a general social view." But Levy went further. He saw the crisis in theoretical physics as fundamentally an aspect of the whole crisis in bourgeois economy and culture. He generalized:

> Any scientific theory is necessarily the specialized development of a general social view, even although those who initiate the theory may be profoundly unaware of the connection. … Just as soon as the categories of social life begin themselves to shift, as in the present, so also, therefore, will a movement of a similar nature be reflected within the inner structure of theoretical science. A crisis in Society will reflect itself indeed in a crisis in ideology, and in a series of crises in diverse branches of science and art. All theories become the subject of fundamental criticism. … That crisis which is emerging in the realms of Relativity and in Quantum Theory, at the macroscopic and the microscopic levels, is therefore fundamentally a partial aspect of the whole crisis in bourgeois economy.

---

[6] I realize that this is more of a postulate than a demonstrated inference. The cultural climate may have interfered with the physical sciences in subtle ways that will only be uncovered by further and more detailed investigations.

However, the claim is not very convincing. It is true that there was a sense of crisis in parts of fundamental physics in the 1930s, and also that there was an increased willingness to consider drastic or even revolutionary solutions. But the feeling of crisis was largely limited to the problematic quantum electrodynamics and other aspects of quantum mechanics. As far as cosmology and cosmophysics were concerned, these fields were small and not taken very seriously by the majority of physicists and astronomers. What is more important, there were good scientific reasons for the sense of crisis. In order to explain it we do not need to refer to social and ideological developments. Of course, this is not a claim that such developments never affect the cognitive level of the physical sciences, only that they did not play a significant role in the period here considered.

The development of fundamental physics in the decade after 1928 shows similarities as well as dissimilarities with the crisis in atomic physics in the early 1920s, such as analyzed by Forman, Seth and others (Forman 1971; Seth 2007). One specific element that cannot be found in the earlier period is the attempts to formulate a cosmophysics based on an integration of microphysics with cosmology, a line of research pursued in different ways by eminent scientists such as Eddington, Jordan, and Dirac. There was also the difference that many of the leading theoretical physicists of the 1930s had participated in the developments that led to the emergence of quantum mechanics in the years 1925-1926. For this reason they tended to see the new problems of physics from the perspective of the recent history of their science. As exemplified by Heisenberg's letter to Pauli of 1935 (quoted in section 2), some physicists made direct comparisons between the situation in the period 1922-1924 and the one a decade later. On the other hand, although the earlier history sometimes encouraged radical proposals and talks of crisis, no one seriously suggested to get rid of quantum mechanics and start on a fresh, as if quantum mechanics had been a mistake.

As I have illustrated, fundamental physics in the 1930s was characterized by a search for the extreme, both when it came to subjects and to methods. I am much less confident of how "modern" it was, or if the physics of the decade can be understood as part of a modernist trend. In fact, one may argue that the rationalism and deductivism of the cosmophysicists was far from modern, as it was merely Platonism and Pythagoreanism on new bottles. But if this was the case it goes nicely with the political ideologies of the decade. For these, too, were basically old-fashioned. Dressed as they were in modern clothes – some red, some black, and others brown – they were nonetheless variations over themes which go way back in time.